# Ultra-high-Q UV microring resonators based on single-crystalline AlN platform

Xianwen Liu,[1] Alexander W. Bruch,[1] Zheng Gong,[1] Juanjuan Lu,[1] Joshua B. Surya,[1] Liang Zhang,[2] Junxi Wang,[2] Jianchang Yan,[2] and Hong X. Tang[1,*]

[1] *Department of Electrical Engineering, Yale University, New Haven, CT 06511, USA*
[2] *R & D Center for Semiconductor Lighting, Institute of Semiconductors, Chinese Academy of Sciences, Beijing 100083, China*

Development of low-loss photonic components in the ultraviolet (UV) band will open new prospects for classical and quantum optics. Compared with other integrated platforms, aluminum nitride (AlN) is particularly attractive as it features an enormous bandgap of ∼6.2 eV and intrinsic $\chi^{(2)}$ and $\chi^{(3)}$ susceptibilities. In this work, we demonstrate a record quality factor of 2.1 $\times 10^5$ (optical loss ∼ 8 dB/cm) at 390 nm based on single-crystalline AlN microrings. The low-loss AlN UV waveguide represents a significant milestone toward UV photonic integrated circuits as it features full compatibility for future incorporation of AlGaN-based UV emitters and receivers. On-chip UV spectroscopy, nonlinear optics and quantum information processing can also be envisioned.

Integrated photonic components have gained remarkable progress at the telecom band thanks to the maturity of silicon photonics [1]. Nonetheless, extending the operating wavelength to the ultraviolet (UV) range still remains non-trivial, yet is significant for on-chip UV spectroscopy [2], biochemical sensing [3], nonlinear optics, and quantum information processing [4]. Since the relatively narrow bandgap (∼1.1 eV) of silicon limits its utility at short wavelengths, the development of a wide bandgap photonic platform, such as aluminum nitride (AlN), is desirable. It is known that AlN exhibits a large bandgap of ∼6.2 eV [5], and is thereby transparent to the light with wavelengths ($\lambda$) above 200 nm and free of two-photon absorption above 400 nm. This wideband transparency allows AlN photonic components to interact with ions at UV and visible regions, such as ytterbium ($^{171}Yb^{+}$) at 369.5 nm, nitrogen vacancy (NV) centers in diamond at 637 nm, and rubidium ($^{85}Rb$) at 778.1 nm, as required for on-chip quantum computing [6] and precision optical clocks [7]. Compared with silicon nitride and diamond, the intrinsic $\chi^{(2)}$ susceptibility of AlN also makes it a competing platform for on-chip nonlinear interactions [8–10] and quantum frequency conversion [11, 12].

Although polycrystalline AlN has proved as a viable platform for chip-scale nonlinear optics and optomechanics [13], it suffers from a large waveguide attenuation of 650 dB/cm at 400 nm due to defect-related absorption and scattering [14], thereby hampering its applications at UV wavelengths. Recently, an impressive intrinsic $Q$-factor ($Q_{\rm int}$) of 24 k at 369.5 nm is measured in nano-crystalline AlN microrings via a swept UV laser [15], yet is likely still limited by the waveguide sidewall roughness and the nano-crystalline film morphology. In contrast, single-crystalline AlN epitaxially grown on sapphire exhibits superior film quality, and poises to become a versatile photonic platform at short wavelengths. Compared with other integrated platforms, crystalline AlN shows unprecedented advantages to leverage UV photonic integrated circuits for implementing passive/active optoelectronic functionalities by incorporating Al(Ga)N-based UV emitters and detectors [16]. At the telecom band, a high $Q_{\rm int}$ up to 2.5 $\times 10^6$ has been recorded in crystalline AlN microrings [17], yielding nonlinear photonic devices such as broadband Kerr frequency combs [18] and high-efficiency Raman lasers [19]. Nonetheless, the current $Q$-factors of ∼1.1 k−7.3 k recorded in crystalline AlN-based UV resonators at 310 − 411 nm [20–24] are far from the maximum attainable $Q$-factors as limited by the Rayleigh scattering loss ($\propto \lambda^{-4}$).

In this Letter, we investigate the waveguide loss in the UV band for high-quality single-crystalline AlN film. Based on a microring architecture, we achieve a record $Q_{\rm int}$ of 210 k at 390 nm (optical loss ∼ 8 dB/cm). We also show UV transmittance that covers 1-nm range for both transverse electric (TE) and transverse magnetic (TM) modes. By comparing the $Q$-factors of AlN microrings with different widths, the influence of sidewall scattering is confirmed. A reduced Rayleigh scattering is verified by feeding the chip with 455 nm light, where an improved $Q_{\rm int}$ of 398 k (optical loss ∼ 3.5 dB/cm) is attained.

Figure 1(a) depicts the cross sections and simulated modal profiles of AlN-on-sapphire microrings. The fundamental modes at 390 nm are well confined within the resonators, but become more susceptible to the sidewall scattering at a smaller waveguide width. Due to the narrow sweeping range (∼1 nm) of our UV laser source (limited by the phase-matching bandwidth of second-harmonic conversion), we adopt a relatively large radius of 30 $\mu$m to obtain a small free spectral range (FSR) such that multiple UV resonances can be characterized. Meanwhile, we utilize two cascaded microrings with slightly varied radii to provide resonant features distinguishable from the transmittance background. For effective waveguide-to-microring coupling at the challenging UV band, we employ a weakly tapered gap coupler to wrap the microring at an angle of 20° [25] and optimize the center gap (0.12−0.15 $\mu$m) based on a FIMMPROP software simulation.

In the experiment, approximately 0.5 $\mu$m single-crystalline AlN is grown on *c*-plane sapphire by metal-organic chemical vapor deposition (MOCVD). The microrings and associated feeding waveguides are then defined by an 100 kV electron-beam lithography (EBL) system (Raith EBPG 5000+) with a negative FOx-16 resist. Since the AlN-on-sapphire wafer is highly insulating, we spin 300 nm poly(4-styrenesulfonic acid) (PSSA) on top of the FOx-16 resist and then sputter 10 nm gold to mitigate charging effects. The PSSA is water-soluble and helps remove the gold after EBL writing. To ensure a high-contrast e-beam pattern, which is particularly important for UV resonators with a small coupling gap, we utilize a 25% tetramethylammonium hydroxide (TMAH) developer [26]. Then, the pattern is transferred to AlN with an optimized $Cl_2/BCl_3$/Ar-based inductively coupled plasma (ICP) etching. Finally, the wafer is embedded in $SiO_2$ by plasma-enhanced chemical vapor deposition (PECVD) and is then cleaved to expose waveguide facets.

* hong.tang@yale.edu

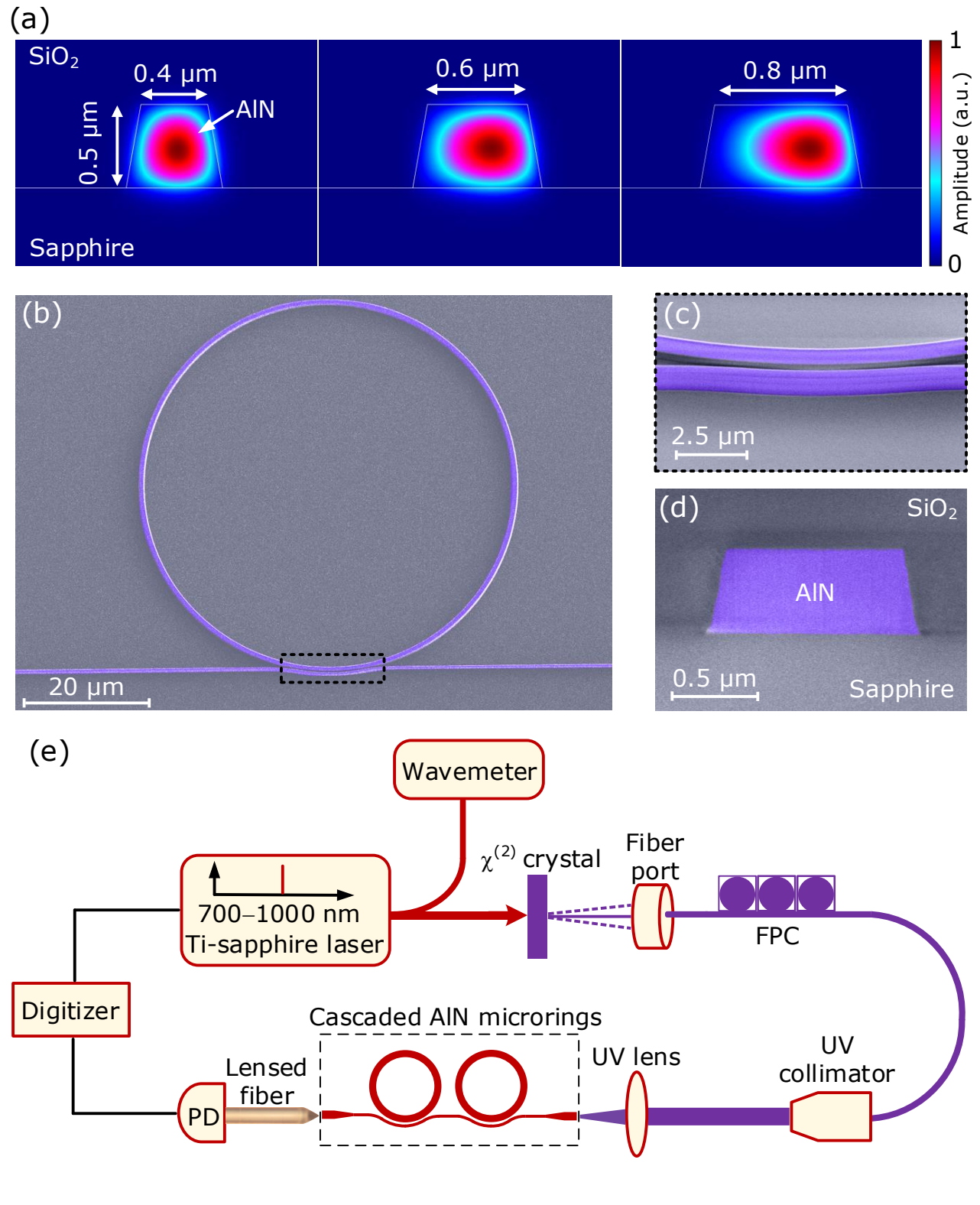

FIG. 1. Device design, fabrication and schematic of measurement setup. (a) Simulated modal profiles of the AlN microrings (top cladding: silicon dioxide, $SiO_2$) at 390 nm by finite element method (FEM). The height and radius of the microring are kept at 0.5 and 30 $\mu$m, while the width is varied from 0.4 to 0.8 $\mu$m. (b)-(d) Colorized SEM images of fabricated devices: (b) top view of the AlN microring without $SiO_2$ cladding (colored purple, the other cascaded ring not shown); (c) zoom in of (b) with tilt view, highlighting a weakly tapered gap coupler with a center separation of 0.14 $\mu$m; (d) cleaved waveguide facet with a $\sim$80$^\circ$ sidewall slope angle and a slight overetching. (e) Schematic of UV transmittance measurement.

Figure 1(b) presents the scanning electron microscope (SEM) image of one of the cascaded AlN microrings. Figure 1(c) is a zoom-in of Fig. 1(b), highlighting a weakly tapered gap coupler (center separation: 0.14 $\mu$m) by tapering the coupling waveguide width from 0.32 $\mu$m to 0.35 $\mu$m when approaching the microring from both sides. To ensure an effective fiber-to-chip coupling for both TE and TM modes, the coupling waveguide is finally tapered to an expanded width of 1 $\mu$m at the chip facets, as shown in Fig. 1(d).

Figure 1(e) illustrates the experimental setup for characterizing the devices. We construct a UV sweeping laser by frequency doubling of a Ti-sapphire laser (M2 SolsTiS, 700−1000 nm) to around 390 nm via a lithium triborate (LBO) $\chi^{(2)}$ crystal. The wavelength of the Ti-sapphire laser is precisely determined by a high-resolution (0.1 pm) wavemeter. Subsequently, we use a UV fiber port to collect the UV beam into a single-mode UV fiber, followed by a UV collimator and a UV lens for coupling the light into the chip. To determine the input polarization, we adjust a fiber polarization controller (FPC) while monitoring the power after a UV linear polarizer (not shown). The output light from the chip is collected by a lensed fiber and then sent to a UV-sensitive photodector (PD). Finally, a data acquisition card is employed to simultaneously record the transmittance from the PD and the wavelength from the Ti-sapphire laser.

It is noteworthy that the generated UV beam exhibits a wavelength-dependent output orientation (indicated by purple dashed lines in Fig. 1(e)) due to the angle-dependent phase matching of the LBO to the pump beam, inducing a varied coupling efficiency of the free-space UV beam into the fiber port upon sweeping the wavelength . For instance, when optimizing the alignment at 390 nm, we observe a large transmission background variation of $\sim$20 dB for the wavelength from 389.5 to 390.5 nm (Supplementary 1). To address this issue, we divide the 1-nm scanning range into several sections, and optimize the alignment separately to suppress the power fluctuation into the fiber.

Figure 2(a) shows the concatenated transmittance of the AlN microring (width: 0.8 $\mu$m; gap: 0.14 $\mu$m) at TM polarization, where a sweeping rate of 2 GHz/s is employed (and hereafter) to ensure a high wavelength resolution. It is noted that the background fluctuation in the $\sim$1 nm full scanning region is suppressed by 13 dB by applying the approach described above. Due to the adoption of two cascaded microrings, we observe four cascaded dips (red and blue lines) spaced by their predicted FSRs, corresponding to the on-resonance fundamental $TM_{00}$ and high-order $TM_{10}$ modes, respectively. According to Figs. 2(b) and 2(c), the measured loaded $Q$-factors ($Q_L$) are 86 k and 103 k for $TM_{00}$ resonances I and II in Fig. 2(a), corresponding to a $Q_{int}$ of 156 k and 189 k at under-coupled condition. The high extinction ratio (ER) of 22 dB at a gap of 0.14 $\mu$m suggests a nearly-critical coupling for the $TM_{00}$ mode. Figure 2(d) shows the transmittance recorded at TE polarization, indicative of cascaded fundamental $\mathrm{TE}_{00}$ resonances with respective FSRs of 657.4 and 657.9 GHz. In Figs. 2(e) and 2(f), the measured $Q_{\mathrm{L}}$ are 144 k and 136 k for $\mathrm{TE}_{00}$ resonances I and II in Fig. 2(d), corresponding to a $Q_{\mathrm{int}}$ of 186 k and 210 k. The highest $Q$-factor at 390 nm attained in this work is a six-fold improvement over the result ($Q_{\mathrm{int}} \sim 35$ k at 400 nm) in nano-crystalline AlN [15].

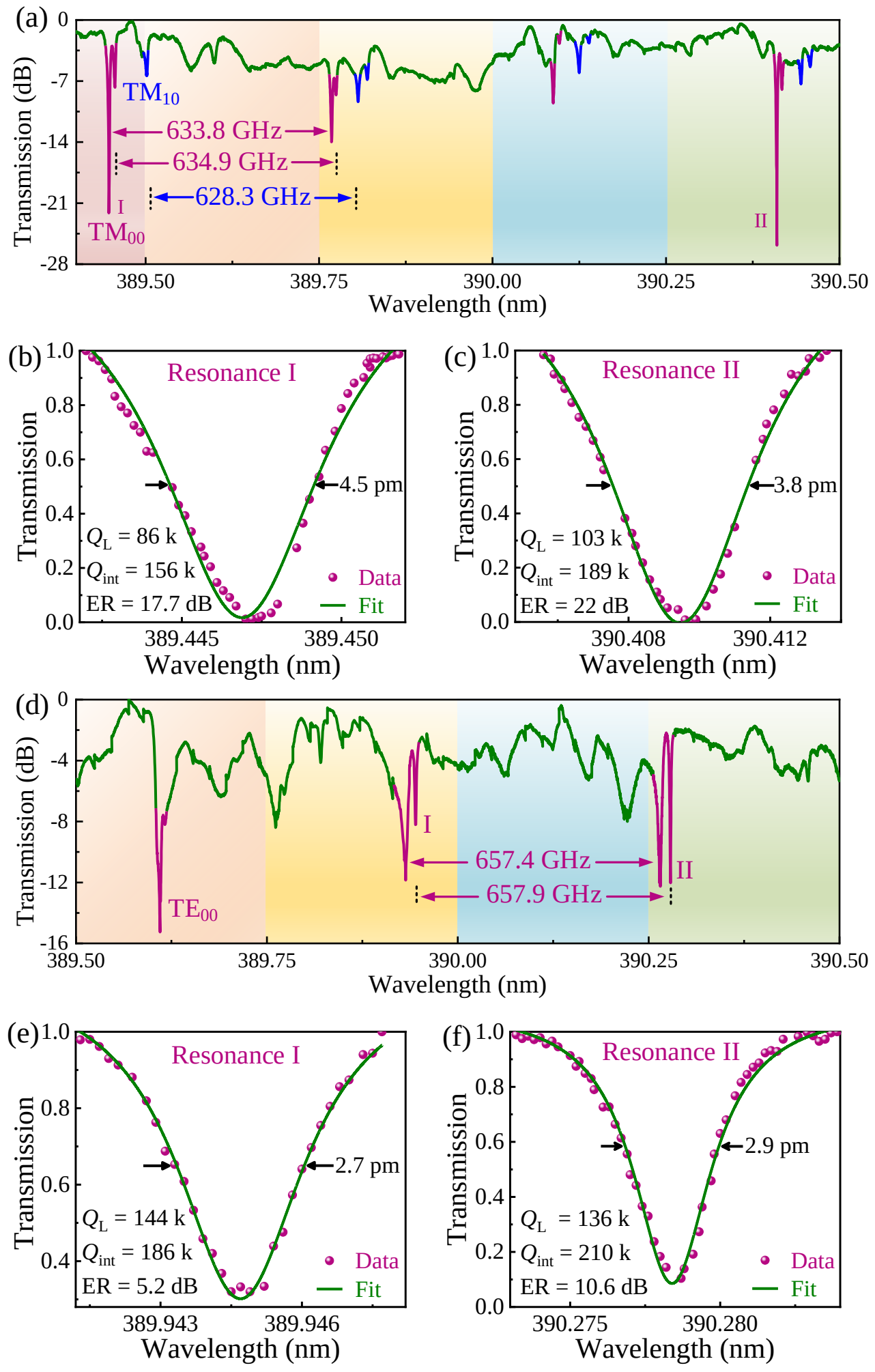

FIG. 2. Characterization of the AlN microring (radius: 30 $\mu$m, cross section: $0.8 \times 0.5$ $\mu$m$^2$, gap: 0.14 $\mu$m) at TM and TE polarizations. The colored regions indicate separated scans with optimal free-space coupling of the UV beam into the fiber port. Here, we extract $Q_{\mathrm{int}} = \frac{2Q_{\mathrm{L}}}{1+\sqrt{\mathrm{T}_0}}$ at under-coupled condition ($\mathrm{T}_0$ being the normalized on-resonance transmission). (a) Microring transmittance showing cascaded resonances for both $\mathrm{TM}_{00}$ and $\mathrm{TM}_{10}$ modes. (b, c) Zoom-in of $\mathrm{TM}_{00}$ resonances I and II in (a), respectively. A Lorentzian fit is applied to each resonance and the large ER indicates nearly-critical coupling. (d) Transmittance with cascaded resonances of $\mathrm{TE}_{00}$ mode. (e, f) Zoom-in of $\mathrm{TE}_{00}$ resonances I and II in (d), respectively.

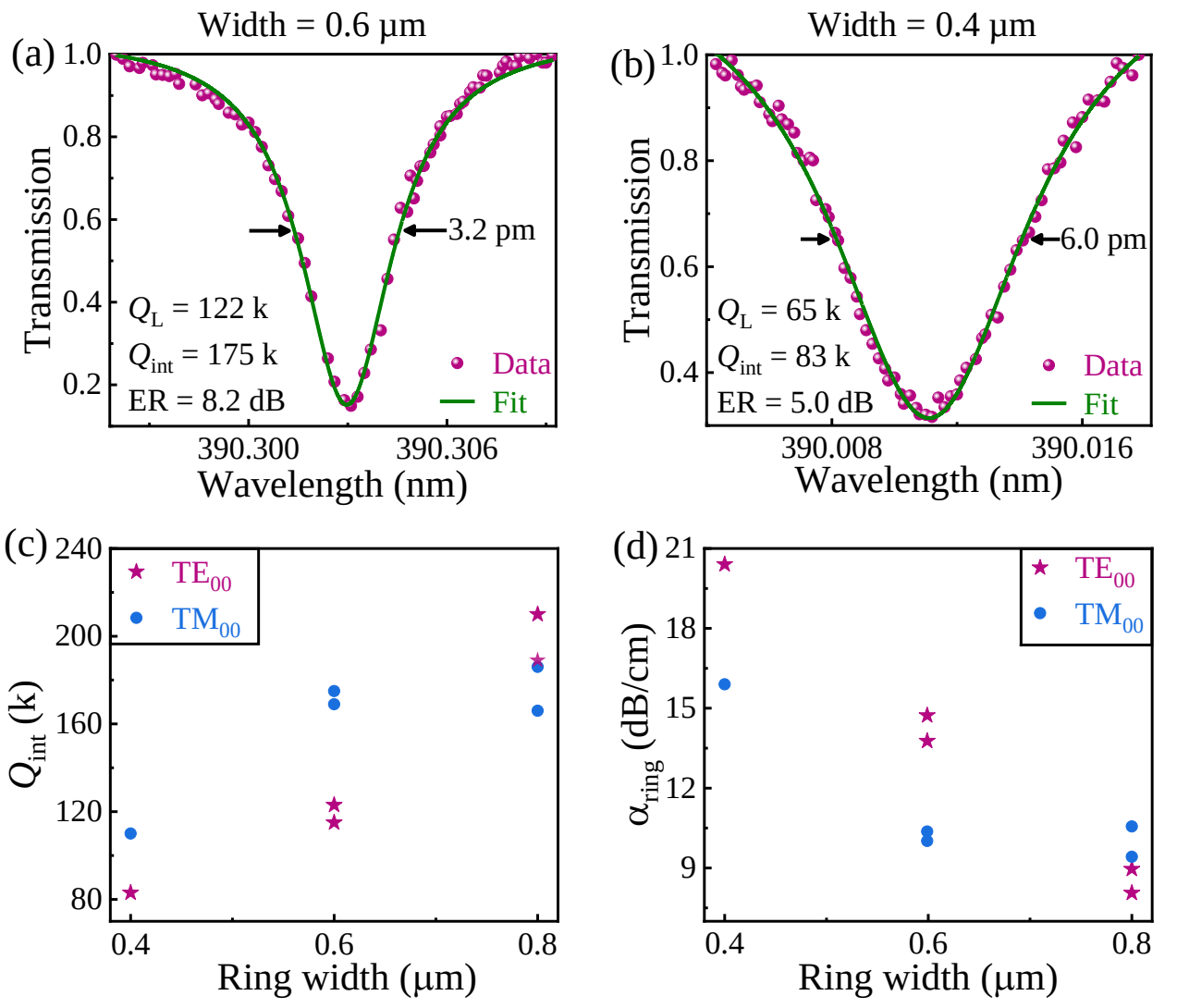

FIG. 3. Comparison of $Q$-factors at 390 nm for devices with different widths (0.4, 0.6, and 0.8 $\mu$m). The $Q_{\mathrm{int}}$ is calculated at under-coupled condition. (a) Measured $\mathrm{TM}_{00}$ resonance of the AlN microring (width: 0.6 $\mu$m; gap: 0.13 $\mu$m) with extracted $Q_{\mathrm{int}}$ of 175 k. (b) Measured $\mathrm{TE}_{00}$ resonance of the AlN microring (width: 0.4 $\mu$m; gap: 0.12 $\mu$m) with extracted $Q_{\mathrm{int}}$ of 83 k. (c) and (d) $Q_{\mathrm{int}}$ and corresponding $\alpha_{\mathrm{ring}}$ versus the width for both $\mathrm{TE}_{00}$ and $\mathrm{TM}_{00}$ modes.

The propagation loss $\alpha_{\mathrm{ring}}$ is then derived to be $\sim$8.0 dB/cm for $\mathrm{TE}_{00}$ mode ($Q_{\mathrm{int}} \sim 210$ k ) based on the expression [17]: $\alpha_{\mathrm{ring}} = 4.343 \times \frac{f_0}{Q_{\mathrm{int}} \cdot \mathrm{FSR} \cdot R_{\mathrm{ring}}}$ ($f_0$ and $R_{\mathrm{ring}}$ being the optical frequency and microring radius, respectively). This value is significantly smaller than the results of 75 dB/cm at 369.5 nm for nano-crystalline AlN microrings [15] and 650 dB/cm at 400 nm for polycrystalline AlN straight waveguides [14]. We believe that the low UV propagation loss achieved in this work is attributed to the excellent film quality of single-crystalline AlN, the engineered microring geometries and the optimized fabrication process. The maximum $Q$-factors that can be achieved in our AlN chips are found to be limited by the sidewall roughness and the Rayleigh scattering, as described later.

We assess the sidewall induced waveguide loss by comparing $Q$-factors of the devices with different widths (0.4, 0.6 and 0.8 $\mu$m). Figure 3(a) shows the measured resonance of $\mathrm{TM}_{00}$ mode for the 0.6-$\mu$m-wide microring. The $Q_{\mathrm{L}}$ is measured to be 122 k, corresponding to a $Q_{\mathrm{int}}$ of 175 k, which is slightly lower than the value in Fig. 2 for the 0.8-$\mu$m-wide device. Nonetheless, a notably reduced $Q_{\mathrm{int}}$ of 83 k is observed for $\mathrm{TE}_{00}$ mode of the 0.4-$\mu$m-wide microring [Fig. 3(b)]. We then plot the extracted $Q_{\mathrm{int}}$ versus the ring width for both $\mathrm{TE}_{00}$ and $\mathrm{TM}_{00}$ modes in Fig. 3(c), where a degraded $Q$-factor with the ring width is clearly observed. We further plot the corresponding $\alpha_{\mathrm{ring}}$ versus the ring width in Fig. 3(d). Notwithstanding a small ring width of 0.4 $\mu$m, we still attain a low $\alpha_{\mathrm{ring}}$ of $\sim$20.4 and $\sim$15.9 dB/cm for $TE_{00}$ and $TM_{00}$ modes, respectively. The experimental results suggest that the sidewall scattering loss $\alpha_{\rm scatter}$ at 390 nm can be mitigated with a relatively large microring width in our AlN material platform.

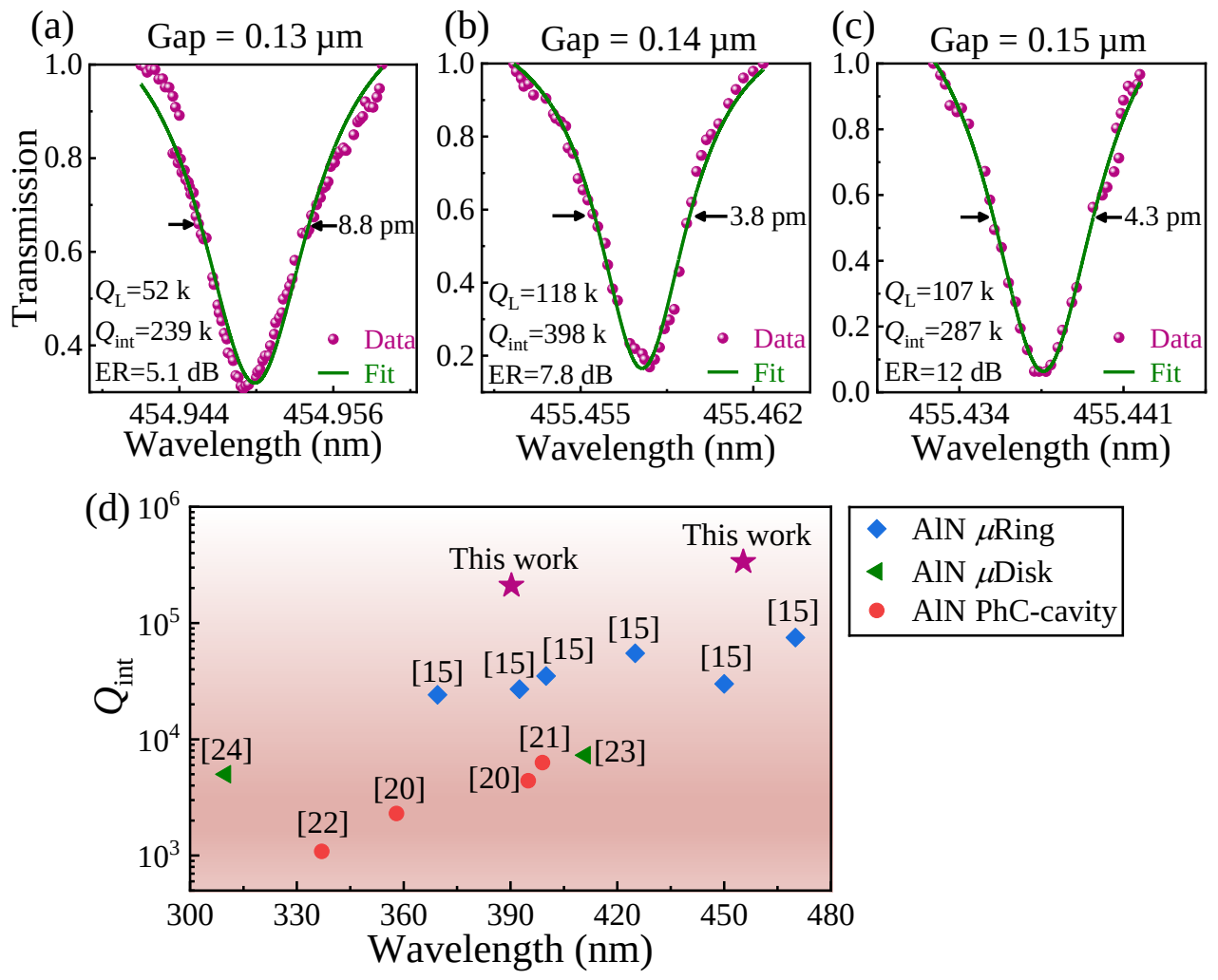

FIG. 4. Measured $Q$-factors of the AlN microrings (width: 0.8 $\mu$m) at 455 nm. (a-c) Resonances of $TE_{00}$ mode at the gap of 0.13, 0.14, and 0.15 $\mu$m, respectively. Here, $Q_{\rm int} = \frac{2Q_{\rm L}}{1-\sqrt{T_0}}$ is extracted at over-coupled condition, as confirmed by the increasing ER with the gap. (d) Comparison of $Q_{\rm int}$ versus the wavelength (below 480 nm) for different microcavities. PhC: photonic crystal; $\mu$Disk: microdisk; $\mu$Ring: microring.

Theoretically, $\alpha_{\rm scatter}$ in optical waveguides is given by [27]:

$$\alpha_{\rm scatter} \propto \frac{\sigma^2 \Delta\varepsilon}{\lambda \cdot n_{\rm eff}} \frac{E_{\rm s}^2}{\iint E^2 {\rm ds}} \quad (1)$$

where $\sigma$ and $n_{\rm eff}$ are the sidewall roughness and effective index, respectively. Meanwhile, $\Delta\varepsilon$ describes the difference of the dielectric constants in the core and cladding. When other parameters are fixed, $\alpha_{\rm scatter}$ is proportional to the normalized modal intensity $E_{\rm s}^2/\iint E^2{\rm ds}$ at the core-cladding interface, which is found to increase with the decreased ring width from our simulation (Supplementary 1). This is consistent with the extracted $\alpha_{\rm ring}$ of $TE_{00}$ mode in Fig. 3(d). The $\alpha_{\rm ring}$ of $TM_{00}$ mode tends to follow the same relationship except at a width of 0.8 $\mu$m. The discrepancy can be explained when accounting for an over-coupled condition at this width, which yields a $Q_{\rm int}$ of 226 k in Fig. 2(d), corresponding to $\alpha_{\rm ring}$ of 7.7 dB/cm for $TM_{00}$ mode. It also agrees with the factor that $TM_{00}$ mode is less susceptible to the sidewall roughness than $TE_{00}$ mode (Supplementary 1), thus is supposed to exhibit a lower propagation loss.

We then explore the influence of Rayleigh scattering by characterizing the $Q$-factors of the 0.8-$\mu$m-wide AlN microring with 455 nm light. We employ another LBO $\chi^{(2)}$ crystal in Fig. 1(e) to attain a continuously tunable blue laser (center at 455 nm), and determine the input polarization with a linear polarizer that covers this region. Figures 4(a-c) depict the measured resonances of $TE_{00}$ mode at a gap of 0.13, 0.14, and 0.15 $\mu$m, respectively. The increasing ER with the gap indicates an over-coupled microring at 455 nm. At a gap of 0.14 $\mu$m, the extracted $Q_{\rm int}$ of 398 k in Fig. 4(b) shows two-fold higher than that at 390 nm [Figs. 2(e) and 2(f)], corresponding to a $\alpha_{\rm ring}$ of $\sim$3.5 dB/cm based on the recorded FSR of $\sim$675.7 GHz for $TE_{00}$ mode. The high $Q$-factors and low propagation loss of AlN resonators at 455 nm are also the state of the art up to date. By including the $Q$-factors attained from our visible and near-infrared single-crystalline AlN microrings, we derive an appropriate $\alpha_{\rm ring} \propto \lambda^{-3}$ (Supplementary 1), suggesting a Rayleigh scattering-dominated loss ($\propto \lambda^{-4}$) in the UV and blue AlN microrings (width: 0.8 $\mu$m). In Fig. 4(d), we summarize the reported $Q_{\rm int}$ at the wavelength below 480 nm for several microcavities. Clearly, the $Q_{\rm int}$ of AlN microrings at 390 and 455 nm in this work are greatly improved over literature reported values.

In conclusion, we demonstrate ultra-high-$Q$ UV microrings based on single-crystalline AlN and investigate possible loss mechanisms. We achieve a record $Q_{\rm int}$ of 210 k at 390 nm, corresponding to a low propagation loss of $\sim$8 dB/cm. Except for the sidewall roughness, we attribute Rayleigh scattering to the loss at UV region based on a higher $Q_{\rm int}$ of 398 k ($\sim$3.5 dB/cm) obtained at 455 nm. Our results pave the avenue to leveraging crystalline AlN-based integrated photonic components at UV and blue regions for a variety of nonlinear and quantum photonic applications by taking advantage of its excellent optical property and intrinsic $\chi^{(2)}$ and $\chi^{(3)}$ susceptibilities.

This work is supported by DARPA SCOUT (W31P4Q-15-1-0006). H.X.T. acknowledges support funding from DARPA's ACES programs as part of the Draper-NIST collaboration (HR0011-16-C-0118), an AFOSR MURI grant (FA9550-15-1-0029), a LPS/ARO grant (W911NF-14-1-0563), a NSF EFRI grant (EFMA-1640959) and David and Lucile Packard Foundation. The authors thank Michael Power and Dr. Michael Rooks for assistance in the device fabrication.

# Supplementary material

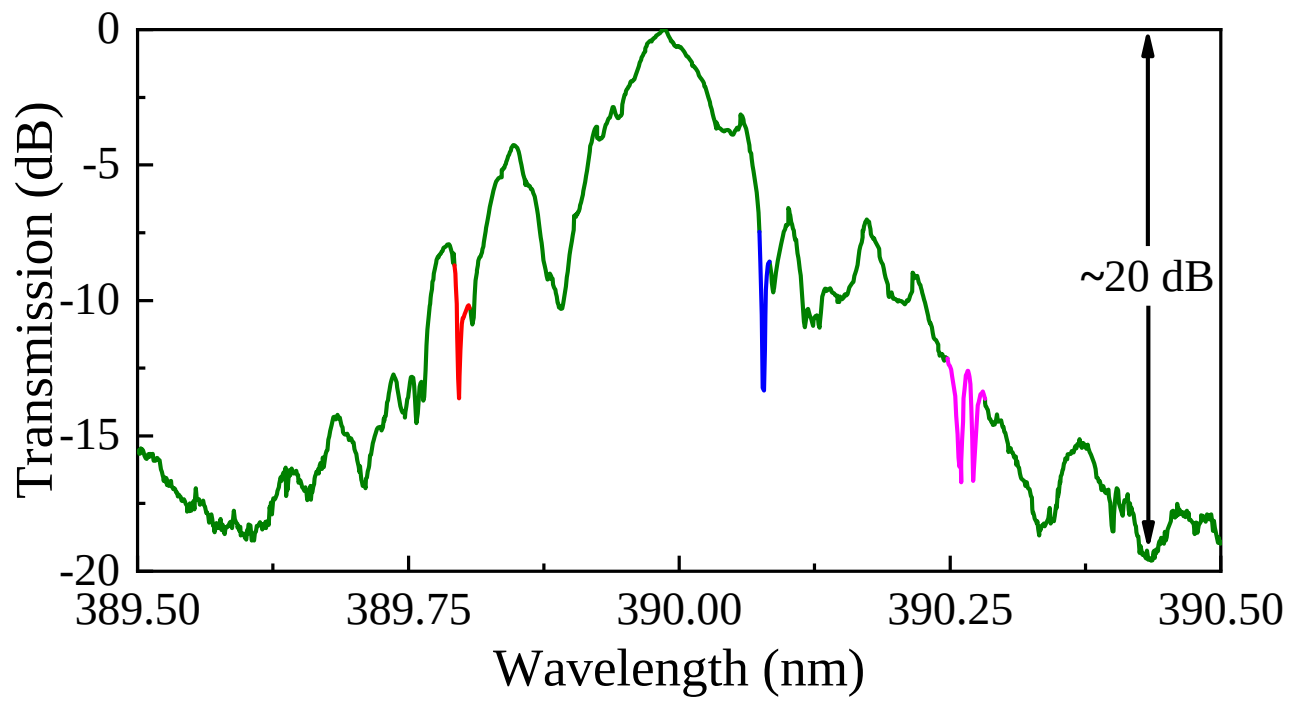

FIG. S1. Transmittance of the aluminum nitride (AlN) microring (width: 0.8 $\mu$m, gap: 0.12 $\mu$m) with the wavelength from 389.5 to 390.5 nm. The coupling of ultraviolet (UV) beam into the following fiber port in Fig. 1(e) of the main text is optimized at 390 nm, which induces a large background fluctuation (green curve) of $\sim$20 dB. The UV resonances are shown in red, blue and pink curves.

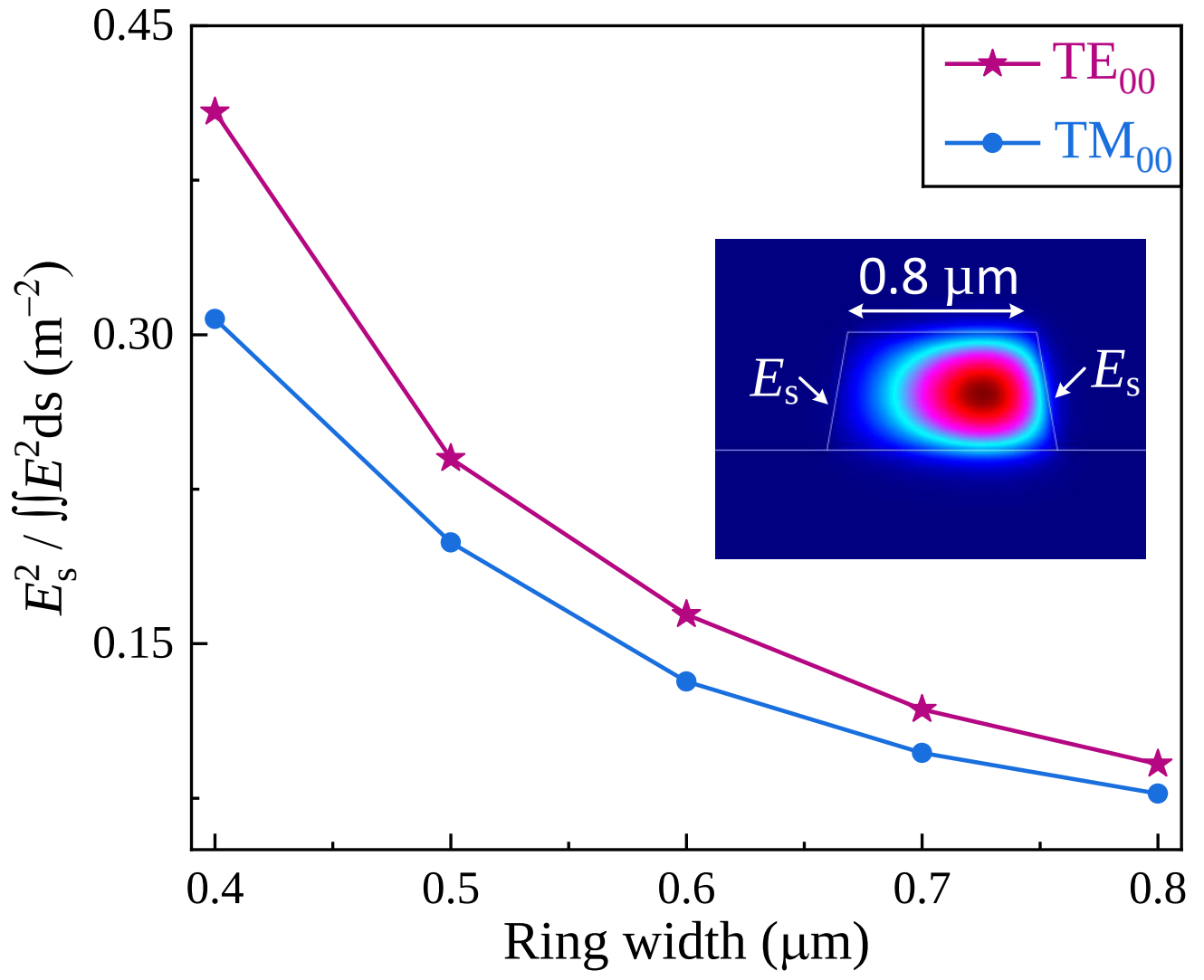

FIG. S2. Simulation of the normalized modal intensity at the core-cladding interface ($E_s^2/\iint E^2$ds) of the AlN microring (radius: 30 $\mu$m) via finite element method. Here, the values of both fundamental transverse electric ($TE_{00}$) and transverse magnetic ($TM_{00}$) modes are calculated. Inset: simulated modal field of 0.8-$\mu$m-wide AlN microring ($E_s$ being the modal field at the waveguide sidewalls). It is evident that a larger $E_s^2/\iint E^2$ds can be identified for the resonator with a smaller width. The $TE_{00}$ mode is also found to exhibit a larger $E_s^2/\iint E^2$ds than that of $TM_{00}$ mode.

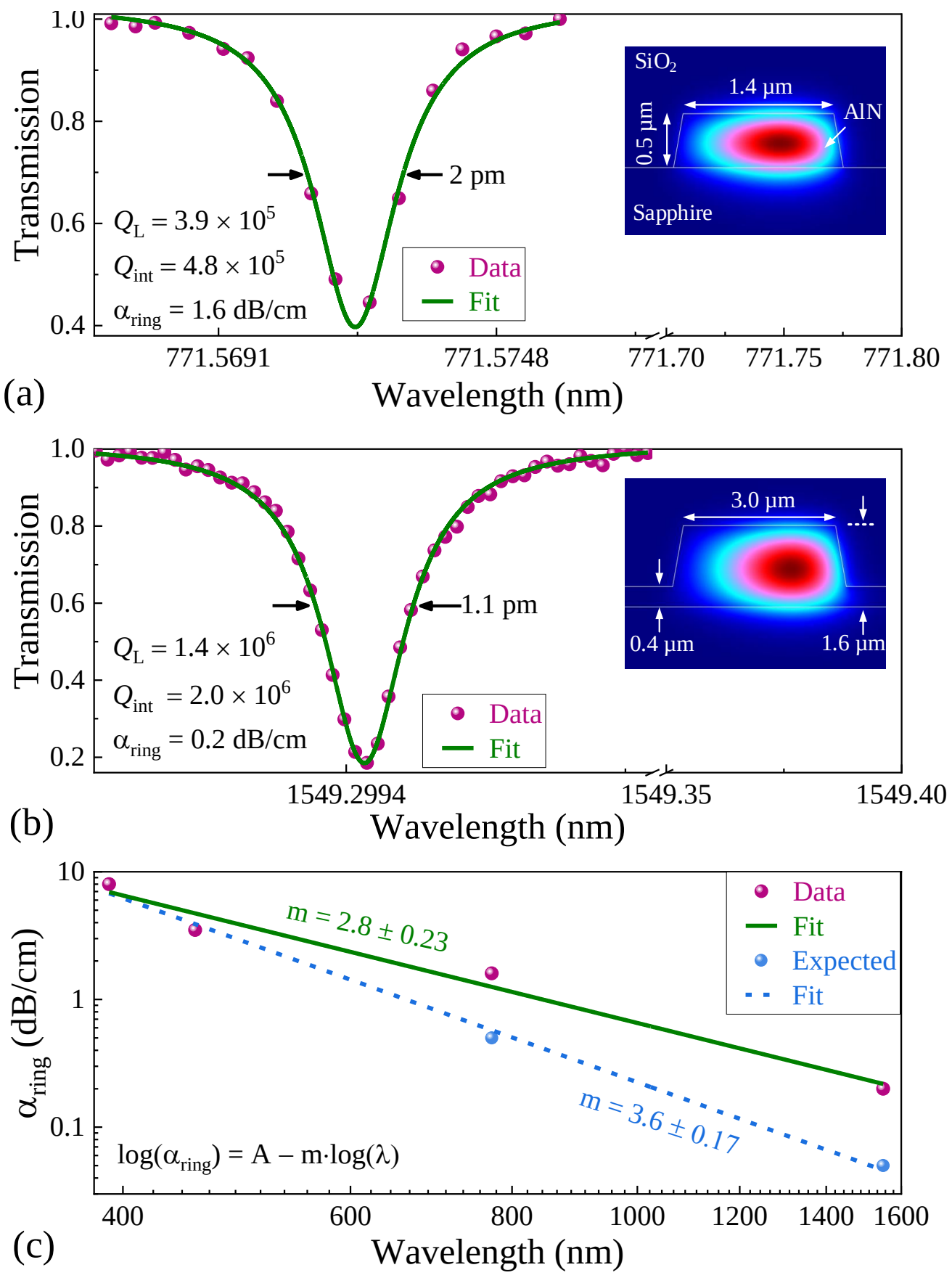

FIG. S3. (a) Measured $TM_{00}$ resonance of a visible AlN microring (radius: 60 $\mu$m, cross section: 1.4 × 0.5 $\mu m^2$, gap: 0.3 $\mu$m) with simulated modal profile in the inset. Here, a Ti-sapphire laser in Fig. 1(e) of the main text is employed to record the data with an extracted $Q_{\rm int}$ of 4.8 × $10^5$ at under-coupled condition, corresponding to a propagation loss ($\alpha_{\rm ring}$) of 1.6 dB/cm. If a thicker AlN film (e.g., 1 $\mu$m) is employed to reduce the scattering loss from the bottom and top interface, a higher visible $Q_{\rm int}$ can be anticipated. (b) Measured $TM_{00}$ resonance of a near-infrared AlN microring (radius: 100 $\mu$m, cross section: 3.0 × 1.6 $\mu m^2$, unetched AlN: 0.4 $\mu$m, gap: 0.6 $\mu$m) via a swept telecom laser. Inset shows the simulated modal profile. The recorded $Q_{\rm int}$ is up to 2.0 × $10^6$ at under-coupled condition, corresponding to a $\alpha_{\rm ring}$ of 0.2 dB/cm. (c) Log-scale plot of $\alpha_{\rm ring}$ recorded in our single-crystalline AlN microrings versus the wavelength, including the $\alpha_{\rm ring}$ of the microring (width: 0.8 $\mu$m) at 390 and 455 nm in the main text. The green linear fit of the experimental data (red) indicates approximately a $\alpha_{\rm ring} \propto \lambda^{-3}$ in our AlN chips. The deviation from the Rayleigh scattering-induced loss ($\propto \lambda^{-4}$) can be explained by excluding the interface scattering loss at visible and near-infrared (NIR) wavelengths, which will give rise to an even lower $\alpha_{\rm ring}$. For instance, by choosing two data points of 0.5 dB/cm and 0.05 dB/cm at visible and NIR regions (light blue plots), we can obtain a $\alpha_{\rm ring} \propto \lambda^{-3.6}$ based on the same data at 390 and 455 nm. This suggests that the $\alpha_{\rm ring}$ at 390 and 455 nm is close to the Rayleigh scattering-induced loss, while the material absorption loss is free in our AlN platform and the scattering loss is mitigated in the AlN microring with a large cross section (0.8 × 0.5 $\mu m^2$).